\documentstyle[preprint,aps,epsfig]{revtex}
\title { One-pion-exchange final-state-interaction and the $p\bar{p} $ near threshold enhancement
in $J/\psi \rightarrow \gamma p\bar{p} $ decays}
\author{B.S.Zou and H.C. Chiang\\
CCAST(World Lab.), P.O.Box 8730, Beijing 100080, China\\
 Institute of High Energy Physics,The Chinese Academy of Sciences,\\
 Beijing 100039, China\footnote{Mailing address}\\
Institute of Theoretical Physics, The Chinese Academy of
Sciences,\\ Beijing 100080, China\\
and\\
Center of Theoretical Nuclear Physics, National Laboratory of
Heavy Ion Accelerator, \\ Lanzhou 730000, China  \\}

\begin{document}
\draft
\maketitle
\date{\today}

\begin{abstract}

For the $N\bar N$ system, the one-pion-exchange (OPE) interaction
gives the largest attractive force for $N\bar N$ with isospin
$I=0$ and spin $S=0$, while a near threshold enhancement was
observed for $p\bar p$ with $I=0$ and $S=0$ in $J/\psi \rightarrow
\gamma p\bar{p} $ decays.  With a K-matrix approach, we find that
the OPE final-state-interaction (FSI) makes an important
contribution to the near-threshold enhancement in the $p\bar{p}$
mass spectrum in $J/\psi \rightarrow \gamma p\bar{p} $ decays.

\end{abstract}
\vskip 2cm

PACS number(s): 14.40.Cs., 13.75.cs., 13.66.Bc.
\newpage
Recently the BES Collaboration has observed  a near-threshold
enhancement in the $p\bar{p}$ invariant mass spectrum from the
radiative decays $J/\psi \rightarrow \gamma p \bar{p}$\cite{BES}.
The enhancement can be fitted with either an S- or P-wave
Breit-Wigner resonance function. No similar structure is seen in
$J/\psi \rightarrow \pi^0 p \bar{p}$ decays. In the S-wave case,
the peak mass is below $2M_p$ around $M=1859 MeV$ with a total
width $\Gamma < 30 MeV$. These observations together with other
similar results in the decays of B mesons\cite{abe} stimulate
further investigations for the quasi-bound nuclear baryonium or
multiquark resonance near the $2M_p$ threshold. What is the origin
of the $p\bar{p}$ enhancement at $M_{p\bar{p}}\approx 2M_p$ in the
radiative decays $J/\psi \rightarrow \gamma p \bar{p}$? Datta et
al.\cite{Dat} describe the enhancement as the formation of a zero
baryon number, ``deuteron-like" singlet $^1S_0$ state. Does it
come from any quasi-bound nuclear baryonium or multiquark
resonance near the $2M_p$ threshold? In order to draw a conclusion
we must study other dynamics which might affect the spectrum of
the outgoing proton and antiproton.

The question of possible nucleon-antinucleon($N\bar{N}$) bound
states was raised many years ago, in particular by Fermi and
Yang\cite{FY}. In the sixties, explicit attempts were made to
describe the spectrum of ordinary mesons as $N\bar{N}$ bound
states. It was noticed \cite{Ball}, however, that the $N\bar{N}$
picture hardly reproduces the observed pattens of the meson
spectrum. Encouraged by the evidence from many intriguing
experimental investigations new types of mesons with a mass near
the $N\bar{N}$ threshold and specific decay properties were
proposed \cite{Sha,Dov}. However, at the time where several
candidate for baryonium were proposed, the quasi-nuclear approach
was seriously challenged by a direct quark picture. Stimulated by
the success of the quark models, exotic multiquark configurations
were studied extensively\cite{Jaf}. The observation of the
Pentaquark state\cite{Penta} stimulates further searches for other
multiquark bound states.

It was noticed\cite{Tornqvist,Ericson,Rich} that for a multiquark
or quasi-bound hadronic system close to its dissociation
threshold, two hadrons will experience their long-range
interaction,in particular the pion exchange. ``Hadronic  molecule"
states might be formed. Due to its long range nature, the pion
exchange plays a crucial role in achieving the binding of some
configuration, especially for two hadrons in relative S-state.  In
a chiral unitary approach it was also found\cite{Oset} that to
solve the coupled channel Bethe-Salpeter equations is crucial for
explaining the observations in the meson-meson and meson-baryon
interactions. Similarly, the outgoing proton and antiproton from
the radiative decays $J/\psi \rightarrow \gamma p \bar{p}$ will
experience the long-range final-state interaction before they are
detected. In order to better understand the nature of the
experimental observation of the near-threshold enhancement in the
$p\bar{p}$ invariant mass spectrum from the radiative decays
$J/\psi \rightarrow \gamma p \bar{p}$ one has to evaluate the
final-state-interaction (FSI) contribution to the invariant mass
spectrum  near $M_{p\bar p}\approx 2M_p$.

  In this note, with the one-pion-exchange (OPE) potential
between the proton and antiproton, we study the FSI of $p\bar{p}$
by the K-matrix approach for the radiative decays $J/\psi
\rightarrow \gamma p \bar{p}$.

It is well known\cite{EW} that for the $NN$ system, the central
OPE potential is attractive for $(S,I)=(0,1)$ (deuteron) or
$(S,I)=(1,0)$, and repulsive for $S=I=0$ (strong) or $S=I=1$
(weak). For the $N\bar N$ system, the meson exchange interaction
is related to the corresponding one for the $N\bar N$ system by
the G-parity transformation, and the OPE potential gets an
additional negative sign due to the negative G-parity of the pion.
Hence the central OPE potential gives the largest attractive force
for $N\bar N$ with $S=I=0$. The attractive force is three times
stronger than the corresponding one for the deuteron. The near
threshold narrow enhancement observed in the $J/\psi \rightarrow
\gamma p \bar{p}$ happens to have quantum numbers $S=I=0$
preferred\cite{BES}. From the one-pion-exchange
theory\cite{EW,Dover}, the nucleon-antinucleon potential can be
written as
\begin{equation}
V_{p\bar{p}}^{\pi}={C^{SI}f^2_\pi\over \vec{q}^2+m_\pi^2}
\end{equation}
with $C^{00}=-3$ for $S=I=0$, $C^{11}=-1/3$ and $C^{10}=C^{01}=1$.
$f_\pi$ is the $\pi NN$ coupling constant $f^2_\pi/4\pi\simeq
0.08$ and $m_{\pi}$ the mass of the $\pi$ meson. $\vec{q}$ is the
3-momentum transfer between the proton and antiproton. The
$p\bar{p}$ with $I=S=L=0$ from the radiative decays $J/\psi
\rightarrow \gamma p \bar{p}$ will experience the largest
attractive long-range OPE final-state-interaction. From the
one-pion-exchange potential, in principle, one could calculate the
two-body $N\bar{N}$ scattering amplitude by solving the
Bethe-Salpeter equation
\begin{equation}
\bar{T}=V+VG\bar{T}.
\end{equation}
Here G is  the loop function of a proton and an antiproton
propagators. It has been shown \cite{Martin} that the K-matrix
formulism provides an elegant way of expressing the unitarity of
the S-matrix for the processes of the type $a+b \rightarrow c+d$.
In the K-matrix approach the invariant S-wave $p\bar{p}$
scattering T-matrix can be expressed as
\begin{equation}
\bar{T}=\frac{K_s}{1-i K_s \rho_{p\bar{p}}}
\end{equation}
where $\rho_{p\bar{p}}$ is the phase space factor for the $p\bar
p$ system
\begin{equation}
\rho_{p\bar{p}}=\frac{M_p^2k}{\pi\sqrt{s}}
\end{equation}
with $s$ the invariant mass squared of the $p\bar{p}$ system and
$k=\sqrt{s/4-M^2_p}$ the momentum magnitude of the proton in the
proton-antiproton c.m. system. Following an usual approach for the
strong interaction in the K-matrix formalism \cite{Zou,Locher},
the $K_s$ is taken as the S-wave projection of the $N\bar{N}$
potential, {\sl i.e.},
\begin{equation}
K_s=\frac{1}{4k^2}\int_{-4k^2}^0 dt V_{p\bar{p}}^{\pi}(t)
\end{equation}
where $t=-\vec{q}^2$. For the $I=S=0$ case, $K_s$ can be easily
evaluated from Eq.(5) as
\begin{equation}
K_s=-\frac{3f^2_\pi}{4k^2}ln(1+\frac{4k^2}{m_{\pi}^2}).
\end{equation}

 In this approach, by considering the OPE FSI of the proton and
 antiproton, the T-matrix for $J/\psi \rightarrow \gamma p \bar{p} (^1S_0)$
decays can be written as
\begin{equation}
T_{J/\psi \rightarrow \gamma p \bar{p}}=\frac{T^{(0)}_{J/\psi
\rightarrow \gamma p \bar{p}}}{1-i \rho_{p\bar{p}}K_s}=
\frac{T^{(0)}_{J/\psi \rightarrow \gamma p \bar{p}}}
{1+i\frac{3M_p^2}{k\sqrt{s}}\frac{f_\pi^2}{4\pi}ln(1+\frac{4k^2}{m_{\pi}^2})}.
\end{equation}
Here $T^{(0)}_{J/\psi \rightarrow \gamma p \bar{p}}$  is T-matrix
of the bare $J/\psi \rightarrow \gamma p \bar{p}(^1S_0)$ without
considering the FSI. The conservation of parity and total angular
momentum requires the orbital angular momentum between $\gamma$
and the $p\bar p(^1S_0)$ to be $L=1$, so that the $T^{(0)}$ is
proportional to the momentum of the photon $K_\gamma$ in the
$J/\psi$ rest system, {\sl i.e.},
\begin{equation}
T^{(0)}_{J/\psi \rightarrow \gamma p \bar{p}}=C K_{\gamma}.
\end{equation}
In reality, C should be an $s$-dependent function. Here to
illustrate the OPE FSI effect, we assume C as a constant. In
Fig.(1) we show the T-matrix squared as a function of the
invariant mass of the proton and antiproton for the $J/\psi
\rightarrow \gamma p \bar{p}(^1S_0)$ process. The solid line
corresponds to that with the FSI and the dashed line is that
without the FSI. We find that the final-state-interaction has an
important contribution to the $p\bar{p} $ enhancement near
$M_{p\bar p}=2M_p $ in $J/\psi \rightarrow \gamma p\bar{p}$
decays. Compared with plateau region well above threshold, the OPE
FSI enhancement factor at the $p\bar p$ threshold is larger than
2. The phenomena of a narrow near-threshold peak due to the
t-channel pion exchange is not new. For example, the striking
narrow peak near $p\omega$ threshold in the $\gamma p\to\omega p$
process is found to be produced by the t-channel pion exchange
\cite{omega-p}.

It is well known that there is a very large production of two
gluon system with $J^{PC}=0^{-+}$ below $2M_p$ from the $J/\psi$
radiative decays \cite{Kopke,BDZ,BES1,BES2,BES3,BES4,BES5}. So C
should at least have some broad resonance peaks below $2M_p$,
which have not been well understood. It is quite possible that the
interference of those components plus the narrow OPE FSI structure
could explain the $p\bar p$ near threshold enhancement in the
$J/\psi \rightarrow \gamma p\bar{p}$ process.

In $J/\psi \rightarrow \pi^0 p \bar{p}$ decays, however, because
of the isospin conservation the isospin of the $p \bar{p}$ system
must be 1. The $p \bar{p}$ interaction is either repulsive or one
order of magnitude weaker than for the isoscalar $p\bar p (^1S_0)$
system. One should not find the near-threshold $p \bar{p}$
enhancement. In the decays of B mesons $B^0 \rightarrow D^0
p\bar{p}$ and
  $B^{\pm} \rightarrow K^{\pm} p \bar{p}$,
 the isospin of the $p \bar{p}$ system has isospin 0. The
 enhancement of the low-mass $p \bar{p}$ systems in B decays may also be
 understood by the FSI. The very narrow proton-antiproton atomic states
 observed by LEAR experiments\cite{atom} at $p\bar p$ threshold
 may also play some role in various narrow structure observed recently
 near $p\bar p$ threshold.

\vspace{1cm}

 We would like to thank profs. Zhang Zong-Ye, Yu You-wen, Shen Peng-nian for
stimulating discussions. This work is supported in part by the
National Natural Science Foundation of China under grand Nos.
10225525, 10055003, 10075057, 10147208 and 19975053 and by the
Chinese Academy of Sciences under project No. KJCX2-SW-N02.

\begin{figure}[htbp]
\vspace*{-0.0cm}
\begin{center}
\hspace*{-0.0cm} \epsfysize=10cm \epsffile{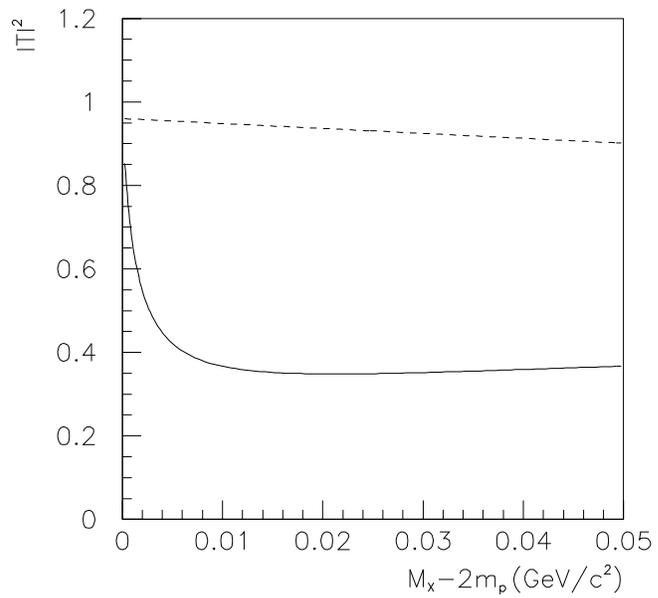}
\end{center}
\caption{T-matrix squared with (solid line) and without (dashed
line) OPE FSI, with an arbitrary normalization. }
\end{figure}

\end{document}